\documentclass[epj,nopacs,referee]{svjour}
\usepackage{graphicx}
\graphicspath{{figures/}}
\usepackage{url}
\usepackage{nicefrac}
\newcommand{\kaos}{{\sc Kaos}\@}

\begin{document}

\title{Exclusive electroproduction of {\boldmath $K^+\!\Lambda$} and
    {\boldmath $K^+\Sigma^0$} final states at 
    {\boldmath $Q^2=$ 0.030--0.055\,(GeV$\!/c$)$^2$}}

\author{
    P.~Achenbach\inst{1}\and 
    C.~Ayerbe Gayoso\inst{1}\and 
    J.C.~Bernauer\inst{1}\fnmsep\inst{6}\and
    S.~Bianchin\inst{2}\and 
    R.~B\"ohm\inst{1}\and
    O.~Borodina\inst{2}\and 
    D.~Bosnar\inst{3}\and 
    M.~B\"osz\inst{1}\and
    V.~Bozkurt\inst{2}\and 
    P.~Byd\v{z}ovsk\'{y}\inst{4}\and 
    L.~Debenjak\inst{5}\and
    M.O.~Distler\inst{1}\and 
    A.~Esser\inst{1}\and 
    I.~Fri\v{s}\v{c}i\'c\inst{3}\and
    M.~{G\'omez Rodr\'iguez}\inst{1}\and
    B.~G\"ok\"uz\"um\inst{2}\and 
    K.~Grie{\ss}inger\inst{1}\and
    P.~Jennewein\inst{1}\and
    E.~Kim\inst{2}\and
    M.~Makek\inst{3}\and 
    H.~Merkel\inst{1}\and
    S.~Minami\inst{2}\and
    U.~M\"uller\inst{1}\and
    D.~Nakajima\inst{2}\and
    L.~Nungesser\inst{1}\and
    B.~\"Ozel-Tashenov\inst{2}\and 
    J.~Pochodzalla\inst{1}\and
    Ch.~Rappold\inst{2}\and
    T.~R.~Saito\inst{1,2}\and 
    S.~{S\'anchez Majos}\inst{1}\and 
    B.S.~Schlimme\inst{1}\and
    S.~\v{S}irca\inst{5}\and 
    M.~Weinriefer\inst{1}\and
    C.J.~Yoon\inst{1}\fnmsep\inst{7}
\newline{A1 Collaboration}
\mail{P. Achenbach, J.-J.-Becherweg 45, D-55099 Mainz, Germany, 
\email{patrick@kph.uni-mainz.de} }
  }
\institute{Institut f\"ur Kernphysik, Johannes
  Gutenberg-Universit\"at, Mainz, Germany \and
  GSI, Helmholtz Center for Heavy Ion Research, Darmstadt,
  Germany \and 
  Department of Physics, University of Zagreb, Croatia \and
  Nuclear Physics Institute, \v{R}e\v{z} near Prague,
  Czech Republic \and
  University of Ljubljana and Jo\v{z}ef Stefan Institute,
  Ljubljana, Slovenia \and
  \emph{Present address:} 
    	MIT-LNS, Cambridge, MA, U.S.A. \and
  \emph{Present address:} 
   Department of Physics and Astronomy, Seoul National University, Korea
   }

\date{Received: date / Revised version: date}

\abstract{
  Cross section measurements of the exclusive
  $p(e,e'K^+)\Lambda,\Sigma^0$ electroproduction reactions have been
  performed at the Mainz Microtron MAMI in the A1 spectrometer
  facility using for the first time the \kaos\ spectrometer for kaon
  detection. These processes were studied in a kinematical region not
  covered by any previous experiment. The nucleon was probed in its
  third resonance region with virtual photons of low four-momenta, $Q^2=$
  0.030--0.055\,(GeV$\!/c$)$^2$. The MAMI data indicate a smooth transition
  in $Q^2$ from photoproduction to electroproduction cross sections.
  Comparison with predictions of effective Lagrangian models based on the
  isobar approach reveal that strong longitudinal couplings of the virtual photon 
  to the $N^*$ resonances can be excluded from these models.
}

\titlerunning{Exclusive electroproduction of $K^+\!\Lambda$ and
    $K^+\Sigma^0$ final states at $Q^2=$ 0.030--0.055\,(GeV$\!/c$)$^2$}
\authorrunning{P. Achenbach et al.}

\maketitle
%


\section{Introduction}

The Mainz Microtron MAMI at the Institut f\"ur Kernphysik in Mainz is
an accelerator to study hadrons with the electromagnetic probe~\cite{Kaiser2008}. 
The exclusive production of mesons by electron and photon beams impinging
on liquid hydrogen targets has been proven to be a valuable tool for
investigating the hadronic structure of the nucleon.  At the energy
scale of the nucleon mass, hadrons are complex systems, whose
description by fundamental equations for the dynamics of
asymptotically free quarks and gluons is complicated by the
non-perturbative nature of QCD. Instead, a successful description of
these reactions has been obtained with hadronic field theories. The
approach is based on effective degrees-of-freedom, where mesons and
baryons are treated as fundamental objects which interact with one
another, characterized by properties such as mass, charge, spin,
parity, form factors, and coupling constants.  Studies of strange
final states provide additional information on the baryonic
resonances.

In one particular type of effective Lagrangian model, commonly
referred to as the isobar approach, the reaction amplitude is constructed
in the lowest-order (tree level) assuming Born terms and exchanges of
various nucleon, hyperon, and meson resonances in the $s$-, $u$-, and
$t$-channels, respectively. Many such models use the single-channel
approach, in which final-state interactions are neglected. Their
copious applications into the strangeness sector started in the 1980s,
e.g.~\cite{Adelseck1990,Williams1992,David1996,Mizutani1998,MartAndMAID2000}. 
For the electromagnetic kaon
production in the so-called third resonance region, many resonances
contribute, which presently cannot be described uniquely by a single
model.

In the one-photon exchange approximation for electromagnetic scattering
of unpolarized electrons off unpolarized target nucleons with coincident
kaon detection 
the virtual photoproduction cross section can be 
expressed as
\begin{eqnarray}
\frac{d\sigma_v}{d\Omega_K^{cm}} & = & \sigma_T + \varepsilon\, \sigma_L +
\sqrt{2 \varepsilon (\varepsilon+1)}\, \sigma_{TL} \cos\phi_K \nonumber\\
 & & +\, \varepsilon\, \sigma_{TT} \cos 2 \phi_K \, ,
\label{eq:xsec}
\end{eqnarray}
where the terms indexed by {\it \scriptsize  T, L, TL, TT} are the transverse,
longitudinal and interference structure functions that 
depend on the virtual photon's four-momentum $Q^2$ and the hadronic energy $W$, and 
$\phi_K$ is the angle between the electron-scattering and 
hadron-production planes. This expression shows the
close connection between electro- and photoproduction. At MAMI the
transitional region at $Q^2 < 0.5$\,(GeV$\!/c$)$^2$ is accessible with
virtual photons that are polarized with a degree-of-polarization $\varepsilon$. 
Longitudinal and longitudinal-transverse 
interference structure functions are accessible only in electroproduction 
experiments. Although not necessarily being small they are suppressed in the total 
cross section through a factor $Q^2/\omega^2$, with $\omega$ being the photon energy.
The $\sigma_{TT}$ and $\sigma_T$ structure functions are related to
the polarized photon asymmetry $\Sigma$, that could be measured in 
photoproduction experiments with real polarized photons, by the relation
$\Sigma = -\sigma_{TT}(Q^2
  \rightarrow 0)/{\sigma_{T}(Q^2 \rightarrow 0)}$. The
transverse-transverse interference structure function for virtual
photons accounts for their partial linear polarization.  With these
connections it is instructive to compare the electroproduction cross
section at very low $Q^2$ directly to unpolarized
photoproduction. For real unpolarized photons 
and unpolarized target nucleons only the transverse structure function remains.

In order to provide a comprehensive understanding of the elementary
kaon production reaction, large kinematic coverage of experimental
data on photo- and electroproduction is needed.  Although recent
electroproduction measurements with high statistics have been performed at Jefferson
Lab~\cite{Mohring2003,Bradford2006,Ambrozewicz2007,Coman2010,McCracken2010}, 
the region of very small momentum transfers was up to now not covered
experimentally. A phenomenological 
extrapolation of the CLAS data from $Q^2 >$ 0.65\,(GeV$\!/c)^2$ to the 
photoproduction point at $Q^2=$ 0 leads to
values of the transverse structure function $\sigma_T$ systematically 
larger than the measured values in photoproduction~\cite{Ambrozewicz2007}.
This unknown transitional region is particularly interesting 
since predictions of models for the separated cross sections differ 
significantly here, see e.g.\ Fig.~3 in~\cite{Coman2010}. 
The goal of the first measurement at MAMI was to
probe the elementary reaction at small momentum transfers and to
determine the angular dependence of the electroproduction cross
section in this kinematic region.

\section{Experiment}

First experiments on the electroproduction of kaons off a liquid
hydrogen target were performed in Mainz with an unpolarized electron
beam of 1.508\,GeV energy in the years 2008--9. The liquid hydrogen 
target cell was 11\,mm wide and 48\,mm long 
with walls made of 10\,$\mu$m thick havar foil. A current of
1--4\,$\mu$A was rastered with a few kHz 
in the transverse directions with an amplitude of $\pm$ 5\,mm
in order to avoid local boiling of the liquid. 
The scattered electrons from the $p(e,e'K^+)$
reaction were detected in SpekB, one of the high-resolution magnetic
spectrometers of the A1 collaboration's spectrometer
facility~\cite{Blomqvist1998}, and the positive kaons in the \kaos\
spectrometer~\cite{Achenbach2011:EPJST}. With its very compact
design and a length of the central trajectory of only 5.3\,m to the 
focal plane, the \kaos\ spectrometer complements the facility in reactions with open
strangeness.

In two of the experimental settings, the central momentum 
for the kaon arm was 0.53\,GeV$\!/c$ and
for the electron arm 0.33\,GeV$\!/c$, respectively 0.45\,GeV$\!/c$. 
The central spectrometer angle
of the kaon arm was 31.5$^\circ$ with a large angular acceptance in
the dispersive plane of $\theta^{lab}_K=$ 21--43$^\circ$. The electron
spectrometer was fixed at the minimum forward angle of
$\theta^{lab}_{e'}\approx$ 15$^\circ$, thereby maximising the virtual photon
flux. The coverage of the photon's four-momenta was $Q^2=$ 0.030--0.055\,(GeV$\!/c$)$^2$  with central values of $Q^2=$ 0.036 and 0.050\,(GeV$\!/c$)$^2$ for the two kinematic settings. The 
corresponding degrees-of-polarization were $\varepsilon=$ 0.4 and 0.54. 
Photon energies, $\omega$, were near the maximum of the kaon production 
cross section at 1.18\,GeV and 1.04\,GeV, 
exciting the hadronic system to invariant energies $W=$
1.75\,GeV and 1.67\,GeV.

Track determination is performed in SpekB by means of two vertical
drift chambers, and timing and trigger signals are provided by two
segmented planes of plastic scintillators immediately behind. A
gas-filled threshold Cherenkov detector is operated to provide a good
separation between pions and electrons. To perform clean electron
identification, a signal in the Cherenkov counter was
required. SpekB reaches a
momentum resolution (FWHM) of $\delta p/p < 10^{-4}$ and an
angular resolution of better than 0.2\,mrad.

In the \kaos\ spectrometer's hadron arm there are two segmented
scintillator walls with 30 paddles each serving as timing, energy
loss, and trigger detectors. Two multi-wire proportional chambers 
serve as coordinate
detectors near the focal plane. A dedicated set of efficiency
counters was built to measure tracking efficiencies for the abundant
pions and protons. The relatively large overlap between the liberated
charges for these two particle species was used to extract the
tracking efficiency for kaons. 

\begin{figure}
  \centering
  \includegraphics[width=0.8\columnwidth]{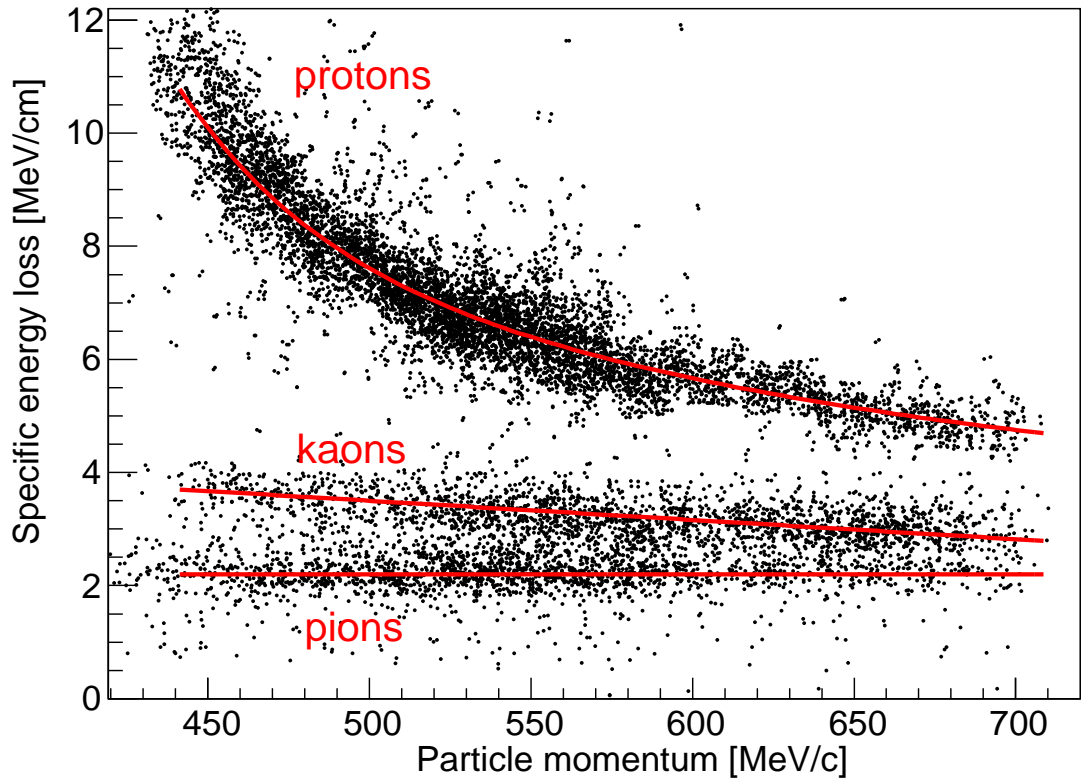}
  \caption{Specific energy loss in one scintillator wall as a function
    of momentum after particle identification cuts. Lines for the
    expected energy losses of pions, kaons, and protons are shown.}
  \label{fig:dEdx}
\end{figure}

\begin{figure}
  \centering
  \includegraphics[width=0.8\columnwidth]{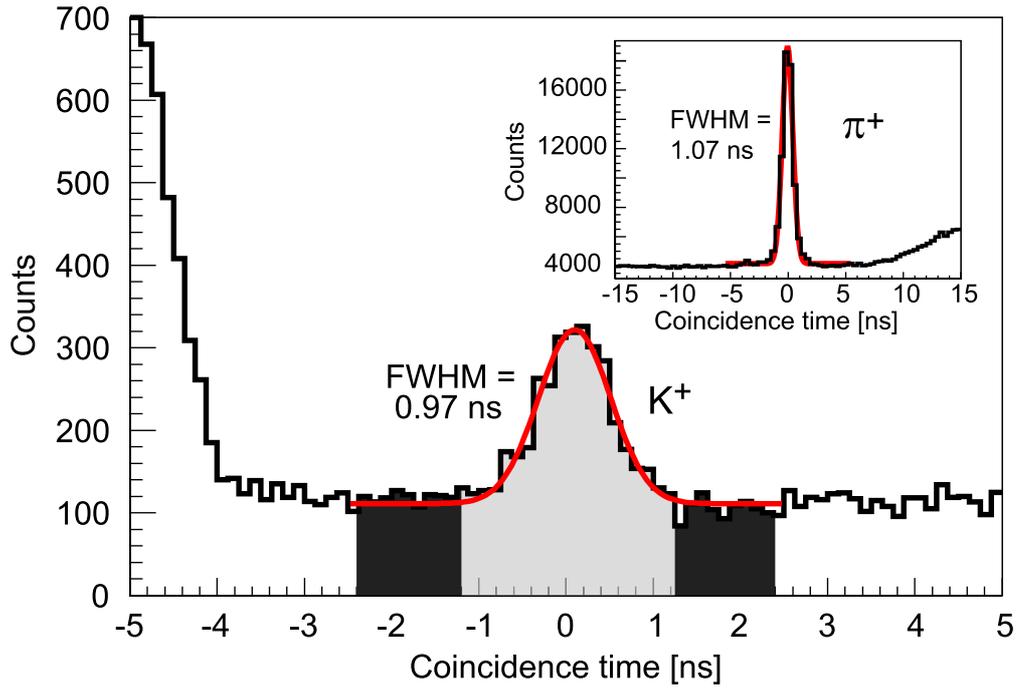}
  \caption{Coincidence time spectra for the $p(e,e'K^+)$ reaction as
    well as for the dominant $p(e,e'\pi^+)$ reaction (insert), after
    corrections for the reconstructed flight path and particle
    identification cuts. Gaussian distributions on top of a constant
    background were fitted to the spectra. The width of the $(e',\pi^+)$
    peak is $\Delta t_{\it FWHM} =$ 1.07\,ns, the width of the
    $(e',K^+)$ peak is $\Delta t_{\it FWHM} =$ 0.97\,ns. The cut regions
    for selecting true and random coincidences of kaons are
    indicated.}
  \label{fig:TOF}
\end{figure}

The intrinsic efficiency of a
single MWPC was measured to be better than 98\,\%. Track
reconstruction efficiencies were dependent on beam intensity, 
being 75--90\,\% at beam
currents of 1--4\,$\mu$A~\cite{Achenbach2011:MWPC}. 
To identify kaons in the measured range of
momenta from 0.4 to 0.7\,GeV$\!/c$, energy loss and flight time were
used.  The signal amplitudes from the individual paddles were
corrected for the reconstructed path length through the scintillator
bulk material and the light absorption inside.  The specific energy
loss corrected for the expected kaon energy loss, $\Delta E^K$, was
required to be within $|\Delta E^K| <$ 640\,keV. See
Fig.~\ref{fig:dEdx} for the measured specific energy loss in one
scintillator wall as a function of momentum after particle
identification cuts. The coincidence time spectra after particle
identification cuts for the $p(e,e'K^+)$ and $p(e,e'\pi^+)$ reactions are
shown in Fig.~\ref{fig:TOF}, where the coincidence time was determined
by using the reconstructed momentum and path length under the
assumption that a kaon or pion was detected. The flight time corrected
for the expected kaon flight time, $\Delta t^K$, was required to be
within $|\Delta t^K| < 1.2$\,ns. The hadron arm trigger was generated
by a combination of the hits in the two scintillator walls.
Since the \kaos\ spectrometer was operated as a single dipole
with open yoke geometry, a study was performed on the contributions 
from coincident particles 
originating outside the spectrometer acceptance that scattered into the 
detectors. These type of events were eliminated by the tracking cuts. 

\begin{figure}
  \centering
  \includegraphics[width=0.8\columnwidth]{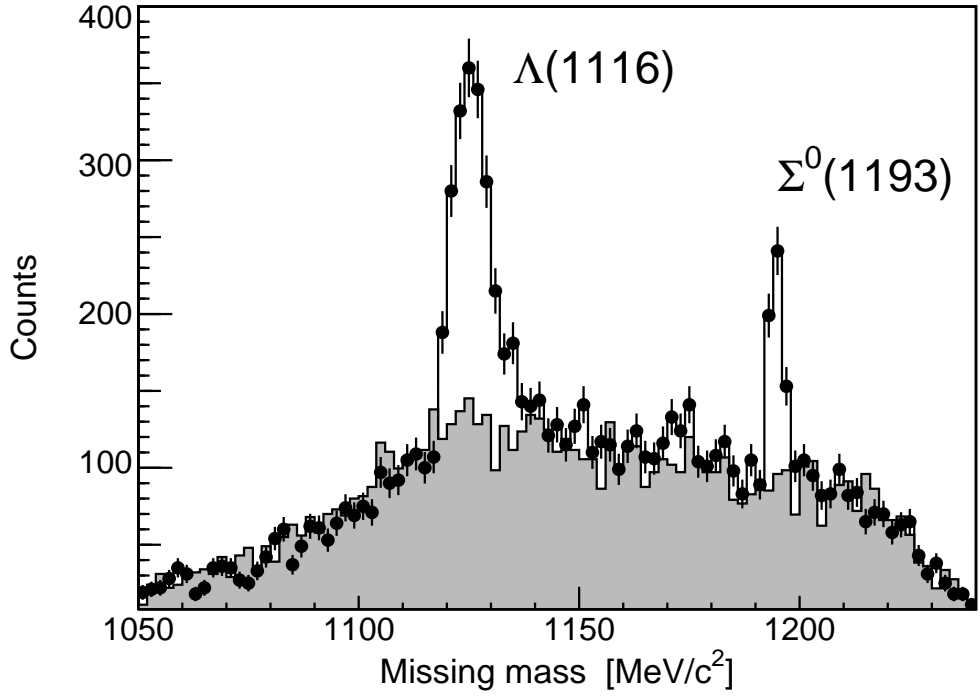}
  \caption{Missing mass spectrum in the $p(e,e'K^+)\Lambda,\Sigma^0$
    reaction for one data set. 
    The shaded histogram shows the missing mass distribution
    in two averaged $(e',K^+)$ coincidence time sidebands with the
    appropriate weights and $\sim$10\% of coincident
    background.}
  \label{fig:MissingMass}
\end{figure}

The measured momenta of the kaon and the electron allow for a full
reconstruction of the missing energy and missing momentum of the
recoiling system.  The missing mass spectrum of one data set is shown in
Fig.~\ref{fig:MissingMass} with the mass resolution being sufficient to
clearly separate $\Lambda$ from $\Sigma$ hyperons as the unobserved baryon
in the reaction. Random background events, identified by two averaged
$(e',K^+)$ coincidence time sidebands and $\sim$10\% of
coincidence background, were subtracted with the appropriate
weights. The mass resolution was limited by the uncertainties in
the transfer matrix. 
For the $\Lambda$ hyperons, events
were selected in the range 1.110 $< M_X$ [GeV$\!/c^2$] $<$ 1.140, 
and for the $\Sigma^0$ hyperons, 
events were selected in the range 1.185 $< M_X$ [GeV$\!/c^2$] $<$ 1.220.

\section{Cross Sections}

The experimental kaon yield, $Y_K$, in the two channels can be related
to the cross section by
\begin{eqnarray}
  Y_K & = & \int\!{\cal L} dt \times \frac{d\sigma_v}{d\Omega_K^{cm}}
  \times \nonumber \\ 
  & & \int\!\! \Gamma(Q^2,W)\, f(Q^2,W)\, A\, R\,
  dV\,,
\end{eqnarray}
where $\Gamma$ is dependent on purely electromagnetic properties and
can be considered as a ``flux of virtual photons", so that
the virtual photoproduction cross section,
$\sigma_v$, could be extracted.
${\cal L}$ is the experimental luminosity that
includes global efficiencies such as dead-times and beam current
dependent corrections such as the tracking efficiency. $A$ is the
acceptance function of the coincidence spectrometer setup, $R$ is the
correction due to radiative and energy losses, and $dV$ the
phase-space element. The accumulated and corrected luminosity in 2009 was 
$\int\! {\cal L}dt \sim$ 2300\,fbarn$^{-1}$. The kaon sample collected in 2008
was significantly smaller. In a Monte Carlo
simulation of the experiment, the phase-space integral was evaluated in
the volume $\Delta V= \Delta Q^2 \Delta W \Delta\phi_K \Delta\Omega_K^{cm}$ 
with limits that extended beyond the physical
acceptances of the spectrometers. 
Radiative corrections in $R$ were included according to~\cite{MoTsai1969}.
The solid angle acceptances for the
electrons and kaons in the laboratory system were $\Omega^{lab}_{e'}
=$ 5.6\,msr and $\Omega^{lab}_K =$ 10.4\,msr.  The two different kaon
momenta associated to $\Lambda$ and $\Sigma^0$ hyperons were
simultaneously within the momentum acceptance of the \kaos\
spectrometer.

The geometrical acceptance of the spectrometer setup, the path length
from the target to the detectors, kaon decay in flight, and kaon multiple 
scattering
were determined using the simulation package {\sf Geant4}.  The kaon
survival fraction varied between 0.2 and 0.35 for the range of momenta
detected.  Fiducial cuts were applied in target acceptance to restrict
events to a region where agreement between the Monte Carlo code and
the analysed data was excellent.

To study the dependence of the measured cross section on the kaon
center-of-mass (cm) angle, $\cos\theta_K^{cm}$, the data was
scaled to the center of the electron-arm 
acceptance.
through the introduction of a scaling function $f(Q^2,W)$ inside
the phase-space integral. 
The scaling minimized the effects of non-uniform
distribution within the $W$ and $Q^2$ bins and the extracted
cross sections can be accurately compared directly to
theoretical calculations at the center of each bin value.
The scaling was performed with the predictions from isobaric models, 
see next Section for their description. The magnitude of the
scaling varied between 0.85 and 1.15 with most of the 
events being scaled by less than 5\%. 
The studied isobar models predicted only small variations of the scaling 
function. 
The model dependence and the uncertainty introduced by the scaling were 
included in the estimate of the systematic uncertainties.

At each kinematic setting, data were partitioned into
several runs. The run-to-run variation of quantities
like dead-time, background particle yields, etc.\ was used as a consistency
check.
The systematic uncertainty assigned to the absolute
cross sections is 8\% and is dominated by the uncertainties
in \kaos\ acceptance (5\%), MWPC tracking efficiencies (4\%), 
analysis cut efficiencies (3\%) and scaling model variation (2\%).
The beam current was measured on
the 1\% level.
Other studied sources of systematic uncertainties included 
kaon survival probability, kaon decay particle mis-identification, effective target length and target 
density fluctuations that introduced a systematic uncertainty
on the percent level.
These uncertainties are significantly smaller than the 
statistical uncertainties in the range of 10--25\%. 

\begin{figure}
  \centering
  \includegraphics[angle=-90,width=0.8\columnwidth]{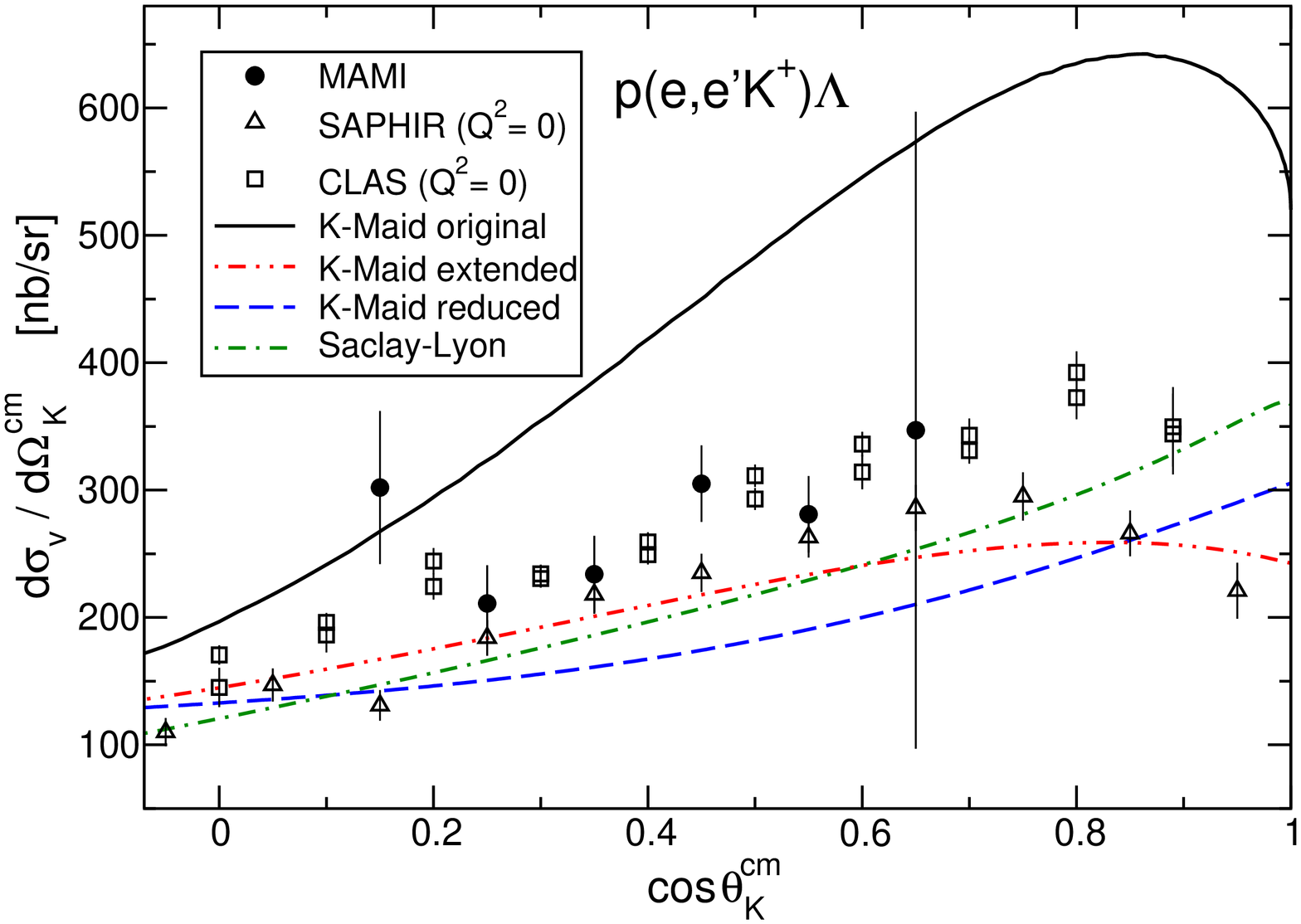}\\[2mm]
  \includegraphics[angle=-90,width=0.8\columnwidth]{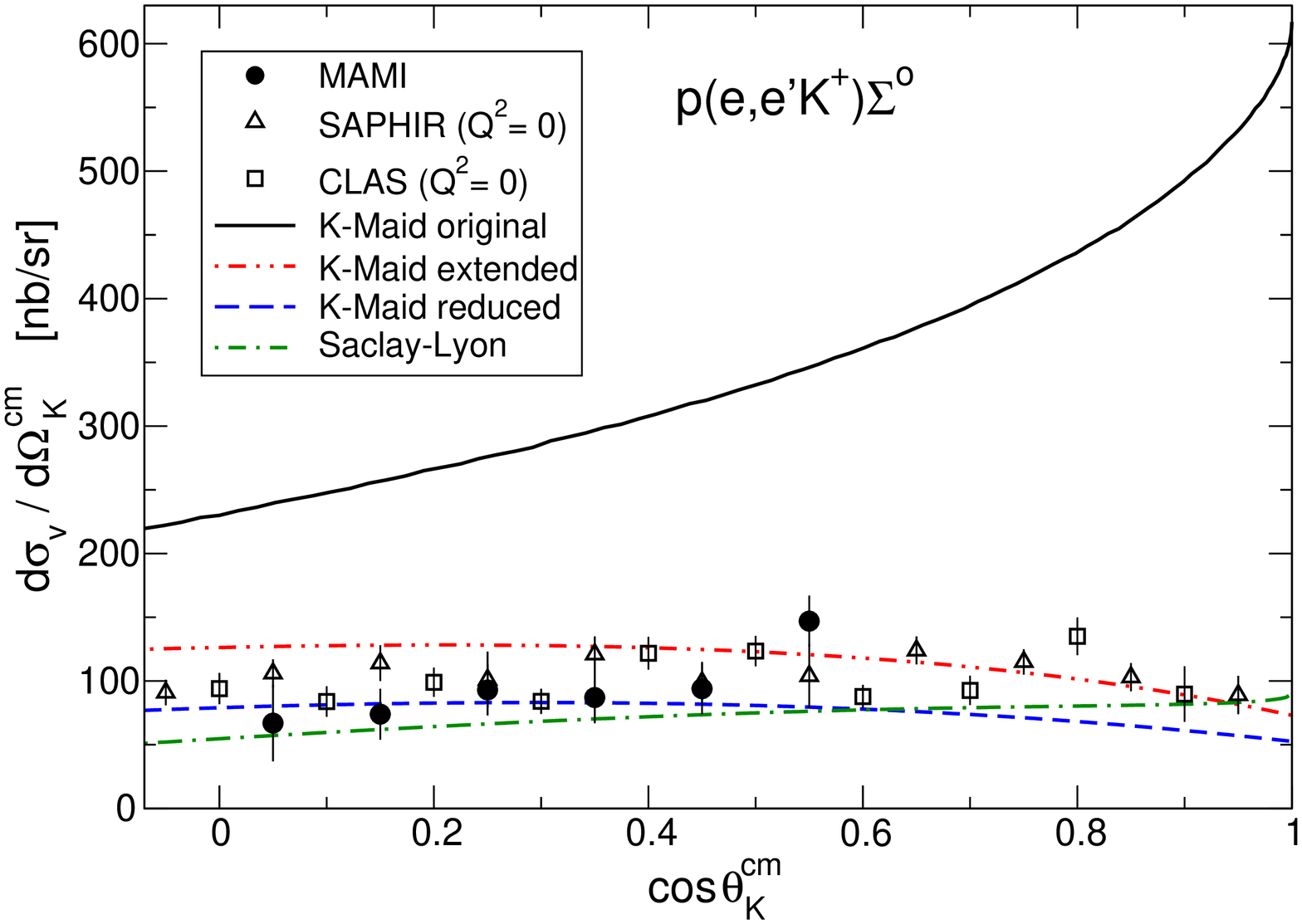}
  \caption{Differential cross sections of kaon electroproduction
    scaled to the center of the experimental acceptance at $\langle
    Q^2 \rangle =$ 0.036\,(GeV$\!/c)^2$, $\langle W \rangle =$
    1.75\,GeV and $\langle \varepsilon \rangle =$ 0.4. The MAMI data
    is compared to variants of the K-Maid
    model~\cite{MartAndMAID2000} (see the text for discussion on the
    variations) and the Saclay-Lyon model~\cite{David1996}. The model
    predictions were averaged in $Q^2$ between
    0.030--0.045\,(GeV$\!/c)^2$ and in $W$ between
    1.74--1.76\,GeV. The photoproduction cross sections at $Q^2=$ 0
    are from the SAPHIR experiment at $W =$
    1.757\,GeV~\cite{Glander2003} and from the CLAS experiment at 
    $W=$ 1.745--1.755\,GeV~\cite{Bradford2006,McCracken2010}.}
  \label{fig:DiffXSec-1750}
\end{figure}
\begin{figure}
  \centering
  \includegraphics[angle=-90,width=0.8\columnwidth]{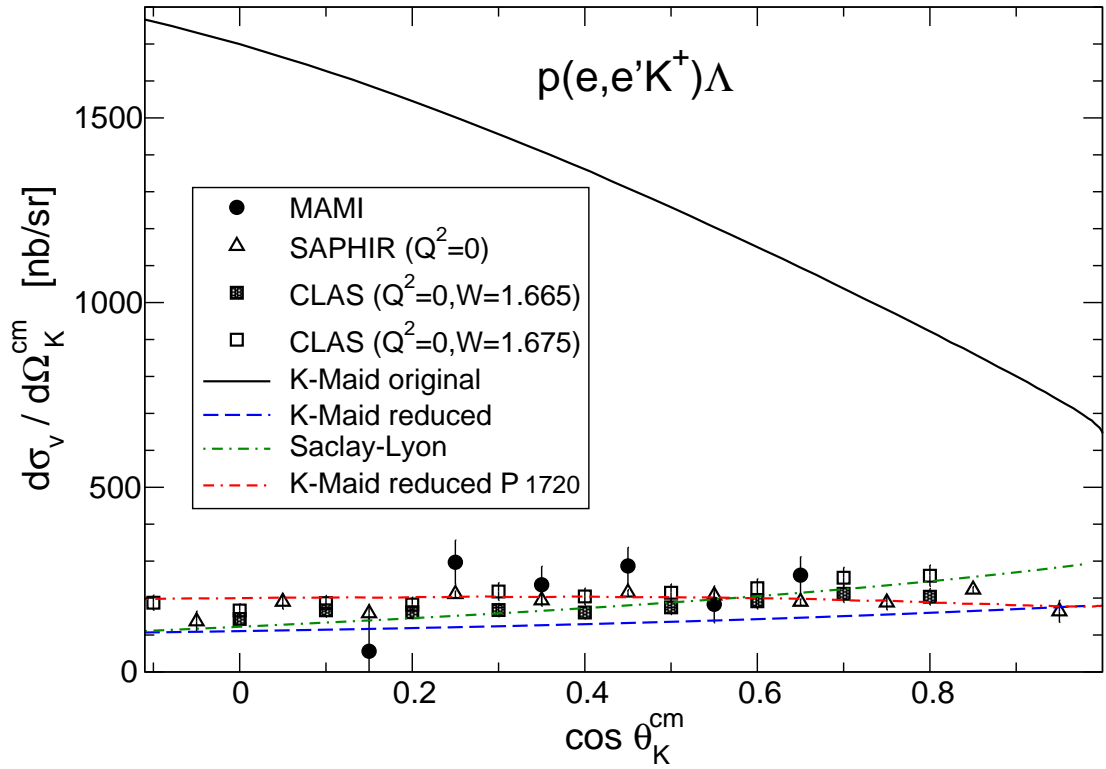}
  \caption{Differential cross sections of kaon
    electroproduction scaled to the center of the experimental acceptance
    at $\langle Q^2 \rangle =$ 0.05\,(GeV$\!/c)^2$,
    $\langle W \rangle =$ 1.67\,GeV and $\langle \varepsilon \rangle
    =$ 0.54. The data is compared to variants of the K-Maid
    model~\cite{MartAndMAID2000} (see the text for discussion on the
    variations) and the Saclay-Lyon model~\cite{David1996}.
    The photoproduction cross sections at $Q^2$ = 0 are from
    the SAPHIR experiment~\cite{Glander2003} and from the CLAS
    experiment at $W=$ 1.665 and
    1.675\,GeV~\cite{Bradford2006,McCracken2010}.}
  \label{fig:DiffXSec-1670}
\end{figure}

The elementary kaon electroproduction cross section was measured at
MAMI in a kinematic region of low $Q^2$ not covered by previous
experiments. The measured angular differential cross sections at $W=$ 1.75\,GeV 
and $\langle Q^2 \rangle=$ 0.036\,(GeV$\!/c)^2$  in the $K^+\Lambda$ and
$K^+\Sigma^0$ reaction channels are shown in Fig.~\ref{fig:DiffXSec-1750}. 
The magnitude of the cross section is comparable to photoproduction data from the
SAPHIR~\cite{Glander2003} and CLAS~\cite{Bradford2006,McCracken2010} 
measurements. Its angular dependence is almost
flat in the $K^+\Sigma^0$ channel whereas in the $K^+\Lambda$ channel 
the data reveals a moderate angular dependence, which is consistent with 
the photoproduction data and with conclusions drawn from the CLAS 
electroproduction data~\cite{Ambrozewicz2007}. In 
Fig.~\ref{fig:DiffXSec-1670} the differential cross section at larger momentum transfer $\langle Q^2 \rangle=$ 0.05\,(GeV$\!/c)^2$ and smaller hadronic energy $\langle W \rangle=$ 1.67\,GeV in the $K\Lambda$ channel is shown.

The MAMI cross section data in the two hyperon channels
is also compared to predictions from variants of
the K-Maid model~\cite{MartAndMAID2000} and from the
Saclay-Lyon model~\cite{David1996}. The K-Maid model includes the kaon resonances
$K^\ast$(890) and $K_1$(1270) in the $t$-channel, as well as four
nucleon resonances, $S_{11}$(1650), $P_{11}$(1710), $P_{13}$(1720),
and a less well known spin-\nicefrac{3}{2} and isospin-\nicefrac{1}{2} resonance near
1.9\,GeV that is found within the context of this model to
contribute in the $s$-channel of the $K^+\Lambda$
production. For $K^+\Sigma$ production the latter is replaced by the
$S_{31}$(1900) and $P_{31}$(1910) $\Delta$-resonances.
Phenomenological form factors are used at the hadron vertices to
account for the high energy region behaviour at $W >$ 1.9\,GeV. No
hyperon resonances are used in this model. Coupling constants are
obliged to fulfill the constraints given by the 20\% broken $SU(3)$ symmetry. 
An interactive version of the
model is available through the internet~\cite{MartAndMAID2000} and is
referred to in this letter as the original variant. 
The parameters of the model, which are necessary to extend the model 
to the electroproduction process, e.g.\ the longitudinal coupling constants,
were fitted using the $Q^2$ dependence between 0.52 and 2.0\,(GeV$\!/c$)$^2$ 
of the longitudinal and transverse cross sections  
measured at Jefferson Lab~\cite{Niculescu1998}. 
Another variant of the model has been constructed, in which
the strong longitudinal couplings to $N^*$ resonances were
removed. This resulted in much smaller longitudinal contributions. 
In the variant labelled P{\tiny 1720} only the longitudinal coupling
to the $P_{13}$(1720) nucleon resonance has been removed. Further, 
some inconsistencies in the
convention for the amplitudes, in the electromagnetic form factors in
the Born terms, and in the couplings to spin-\nicefrac{3}{2} resonances have
been corrected, resulting in minor changes to the predictions. 
This variant is referred to in this letter as the reduced variant. 
Finally, a new version of K-Maid is under construction, that uses very small
longitudinal couplings~\cite{MartAndTiator}.
In this version four nucleon
resonances, $D_{15}(1675)$, $D_{13}(1700)$, $F_{15}(2000)$, and
$D_{15}(2200)$ were added to those assumed in the original version
and the free parameters of the model were re-fitted to describe the new
world data on the photo- and electroproduction of kaons. This version is 
referred to as the extended variant.

The version of the Saclay-Lyon model used in the 
description of the data shares with K-Maid the same kaon resonances
and the $SU(3)$ constraints on the main coupling constants. The set of
nucleon resonances differs from K-Maid and includes resonances with
spins up to \nicefrac{5}{2}, see~\cite{David1996}. 
Instead of hadronic form factors
the spin-\nicefrac{1}{2} hyperon
resonances S$_{01}$(1405), P$_{11}$(1660), S$_{01}$(1670) and
P$_{01}$(1810) are used for
counterbalancing the strength of the Born terms through a destructive
interference with these $u$-channel resonances. For $K^+\Sigma^0$
production, the P$_{33}$(1232), P$_{31}$(1910), and P$_{33}$(1920)
$\Delta$-resonances are added to those given above.
  
\section{Discussion}

The observed behaviour of the cross section 
in the vicinity of the photoproduction point is 
very important for understanding dynamics of the process. 

\begin{figure}
  \centering
  \includegraphics[angle=-90,width=0.8\columnwidth]{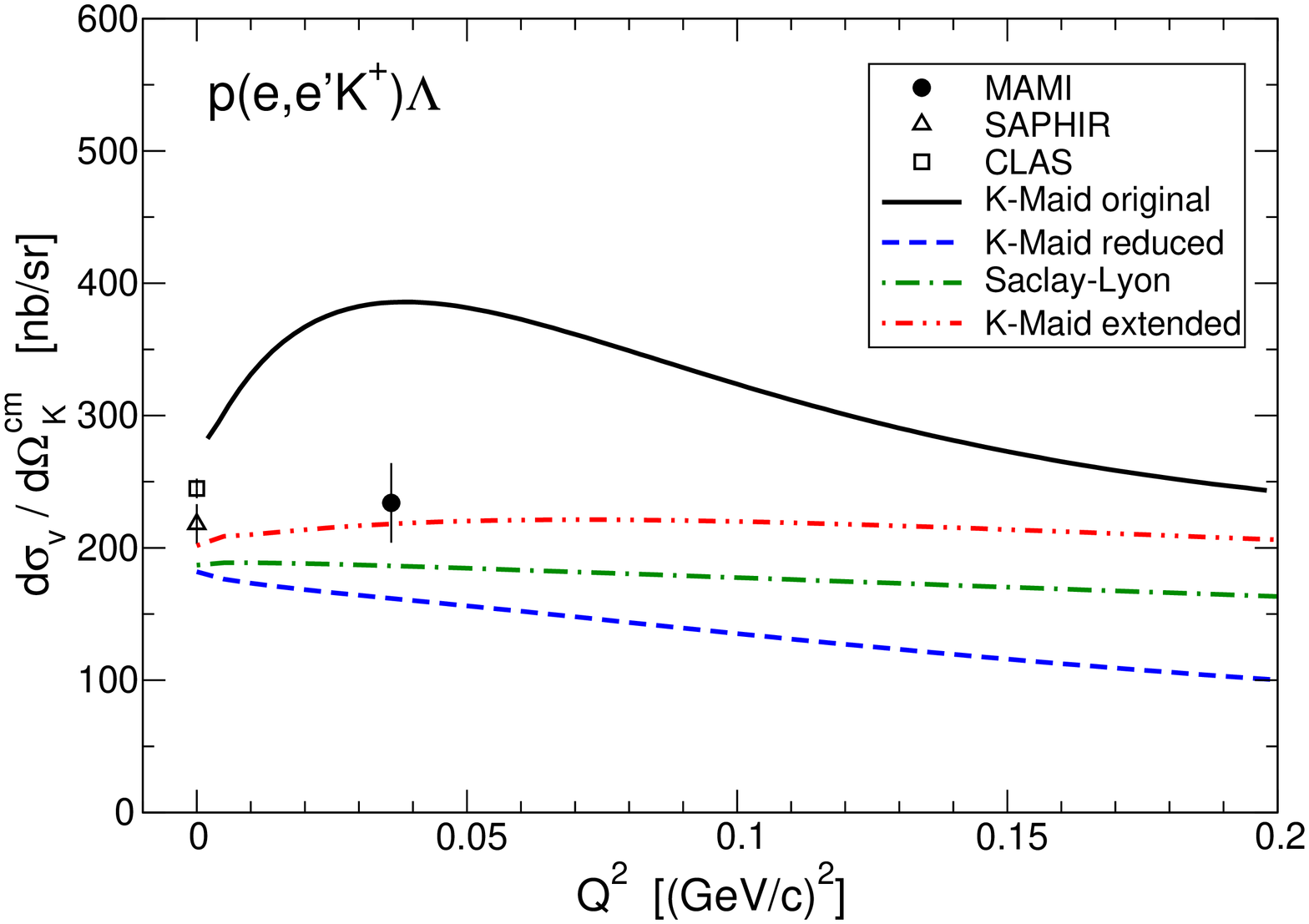}\\[2mm]
  \includegraphics[angle=-90,width=0.8\columnwidth]{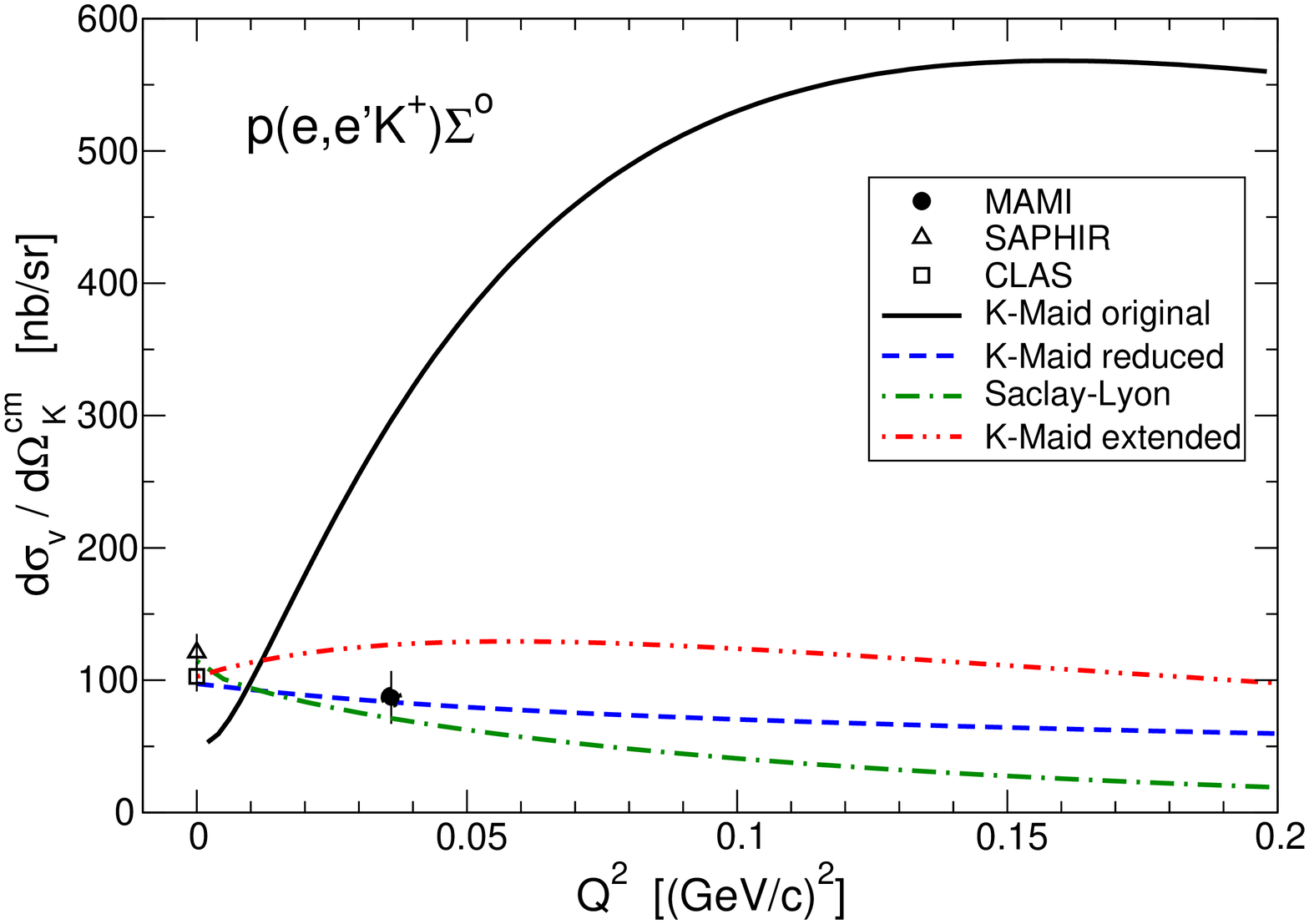}
  \caption{Predictions for the dependence of the kaon electroproduction cross sections 
  	on the virtual photon's four-momentum $Q^2$ at $\cos\theta_K^{cm}=$ 0.35
  	compared to the two experimental 
  	data points from MAMI for the $K^+\Lambda$ and $K^+\Sigma^0$ reaction channels. 
  	References for models and data are the same as for Fig.~\ref{fig:DiffXSec-1750}.}
  \label{fig:Q2dependence}
\end{figure}

The MAMI data indicate a smooth transition 
in $Q^2$ from photoproduction to electroproduction cross sections 
for both reaction channels as shown in Fig.~\ref{fig:Q2dependence}.
The CLAS data on the unpolarized cross section in the $K^+\Lambda$ 
channel, measured for 0.65 $< Q^2 <$ 2.5\,(GeV$\!/c)^2$, 
reveals a fall-off with respect to the photoproduction point~\cite{Ambrozewicz2007}. 
However, the extrapolation of this trend to the photoproduction point 
with a dipole form results in a too large structure function $\sigma_T$~\cite{Ambrozewicz2007}, 
which is apparent especially at forward kaon angles. 
These observations suggest a more complex dynamics of the $K^+\Lambda$ 
electroproduction process in the transition region of 
0 $< Q^2 <$ 0.5\,(GeV$\!/c)^2$. 
Moreover, various models have predicted different transitions
between the photoproduction point and the electroproduction cross
section. 

\begin{figure}
  \centering
  \includegraphics[angle=-90,width=0.8\columnwidth]{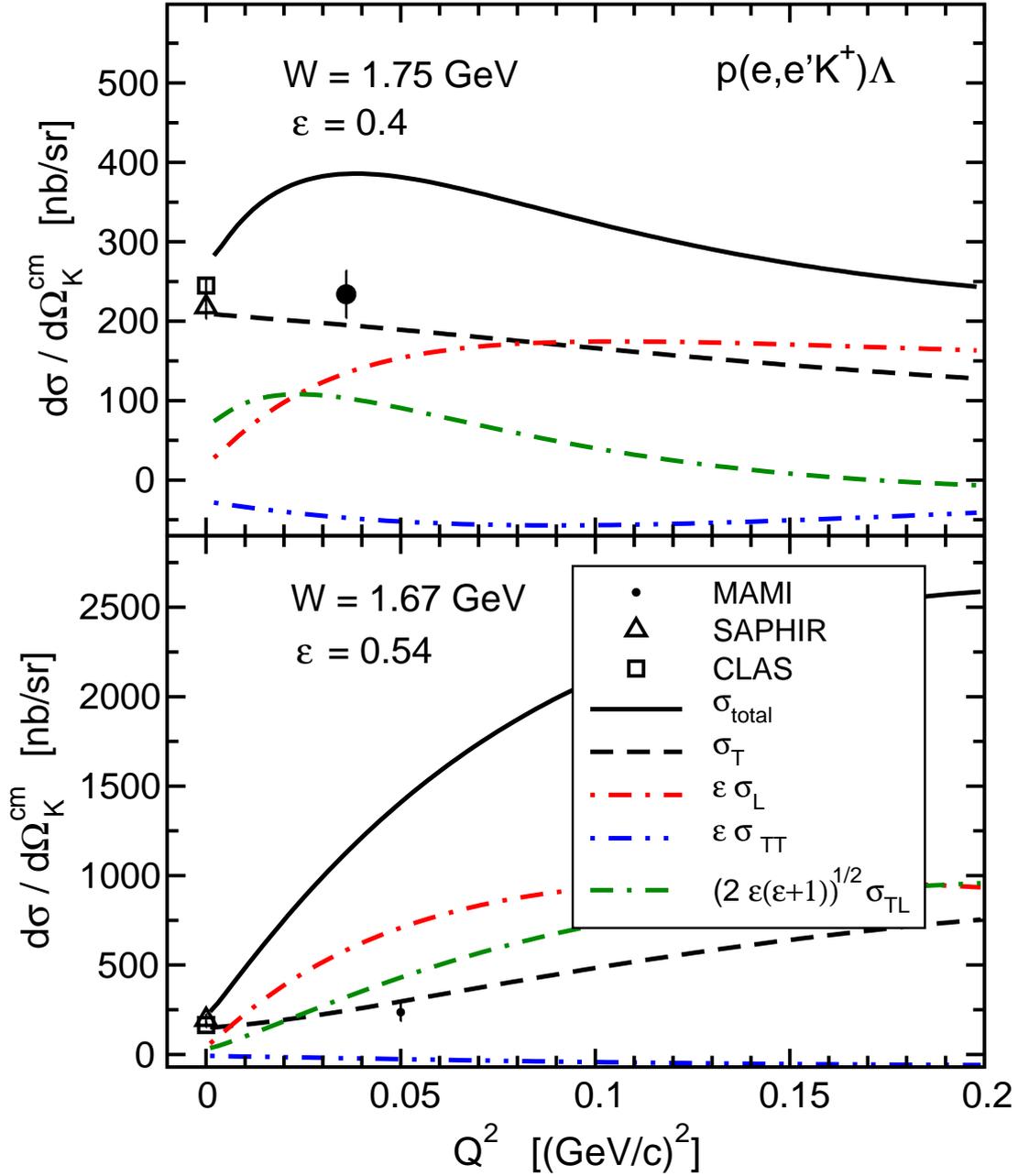}
    \caption{Predictions from the original K-Maid model for the dependence 
    of the kaon electroproduction cross sections on $Q^2$
    at $\cos\theta_K^{cm}=$ 0.35, separated according to Eq.~\ref{eq:xsec}, 
    compared to the data from MAMI for the $K^+\Lambda$ reaction channel. 
    A strong rise with $Q^2$ of the longitudinal, $\sigma_L$, and
    the transverse-longitudinal interference, $\sigma_{TL}$, structure
    functions was predicted by the model,
    especially at $W=$ 1.67\,GeV where the influence of a 
    strong longitudinal coupling of the resonance $P_{13}$(1720) 
    is apparent. In contrast, the transverse structure function, 
    $\sigma_T$, is only slowly varying with $Q^2$.
    References for model and data are the same as for Fig.~\ref{fig:DiffXSec-1750}.}
  \label{fig:responsefunctions}
\end{figure}

Information could also be obtained on the electromagnetic couplings to
hadronic resonances.
In Fig.~\ref{fig:responsefunctions} separated cross sections from the original K-Maid
model are compared to the MAMI data which covers a range in $Q^2$ from 0.030 to 0.055\,(GeV$\!/c$)$^2$. It was predicted
that the longitudinal contribution exceeded the transverse cross
section which was explained in the model by a strong
longitudinal coupling of the electromagnetic field 
to the resonances.  These couplings are given in
the Lagrangian at the nucleon--photon--resonance vertices by a 
term that is proportional to $Q^2$.  
This effect is apparent especially in Fig.~\ref{fig:responsefunctions} 
at $W= $ 1.67\,GeV
where the steep rise of the longitudinal cross sections is due
to the contribution from a strong longitudinal coupling
of the resonance $P_{13}$(1720). In contrast, models
without longitudinal couplings predicted the electroproduction cross
sections below the photoproduction data, which means that in these
variants the total cross section slowly varied with $Q^2$.  
Also, the extended variant of the K-Maid model
reveals a rather flat $Q^2$ dependence of the
cross section in both channels. 

This first open strangeness experiment has also demonstrated that the
\kaos\ spectrometer is, in connection with the high-quality continuous
wave electron beam of MAMI, a very effective tool for investigating
kaon production off nucleons and nuclei with electron scattering.  
Based on the MAMI data some versions of isobar models for the process, which 
assumed strong longitudinal couplings to the nucleon resonances, could be excluded
at the low energies that were probed. Other
versions of isobar models, in which weak or no longitudinal couplings
to $N^*$ resonances appear, predict much smaller total cross sections
and are in better agreement with the data. 
At MAMI, new measurements at very low $Q^2$ with polarized beam will
provide more complete information on the longitudinal-transverse
response in this process and therefore will deliver additional new
constraints on the model parameters. 

To conclude, this measurement and expected new data from MAMI in this
kinematic region will help in better understanding the dynamics of the
electromagnetic kaon production and in the modeling of the nucleon
and its resonances.

\begin{acknowledgement}

We would like to thank the accelerator group of MAMI and the staff of
the workshops for their excellent support. We are thankful to T.~Mart 
from the University of Indonesia for providing the K-Maid code and who 
assisted in the interpretation of the model. We also thank Th.~Walcher
and A.~Jankowiak for their support in the early phase of the project. 

Work supported in part by
the Federal State of Rhineland-Palatinate and by the Deutsche 
Forschungsgemeinschaft 
(DFG) with the Collaborative Research Center 443, 
by the Research Center ``Elementary Forces and Mathematical Foundations'', 
and by the DFG with the HBFG-122-572 grant.  
We also acknowledge
the support by the Research Infrastructure Integrating Activity
``Study of Strongly Interacting Matter´´ HadronPhysics2 under the 7th
Framework Programme of EU 
and by the Grant Agency of the Czech Republic, grant No.~202/08/0984.

\end{acknowledgement}

\bibliographystyle{epj}
\bibliography{references}

\begin{thebibliography}{18}

\bibitem{Kaiser2008}
K.H. Kaiser et~al., Nucl. Instr. Meth. Phys. Res. \textbf{A 593}, 159 (2008)

\bibitem{Adelseck1990}
R.A. Adelseck, B.~Saghai, Phys. Rev. \textbf{C 42}, 108 (1990)

\bibitem{Williams1992}
R.A. Williams, C.R. Ji, S.R. Cotanch, Phys. Rev. \textbf{C 46}, 1617 (1992)

\bibitem{David1996}
J.C. David, C.~Fayard, G.H. Lamot, B.~Saghai, Phys. Rev. \textbf{C 53}, 2613
  (1996)

\bibitem{Mizutani1998}
T.~Mizutani, C.~Fayard, G.H. Lamot, B.~Saghai, Phys. Rev. \textbf{C 58}, 75
  (1998)

\bibitem{MartAndMAID2000}
T.~Mart, C.~Bennhold, Phys. Rev. \textbf{C 61}, 012201 (2000), t.~Mart,
  C.~Bennhold, H.~Haberzettl, L.~Tiator, An effective {Lagrangian} model for
  kaon photo- and electroproduction on the nucleon, online available at
  \url{www.kph.uni-mainz.de/MAID/kaon/kaonmaid.html}

\bibitem{Mohring2003}
R.M. Mohring et~al. (E93-018 Collaboration), Phys. Rev. \textbf{C 67}, 055205
  (2003)

\bibitem{Bradford2006}
R.~Bradford et~al. (CLAS Collaboration), Phys. Rev. \textbf{C 73}, 035202
  (2006)

\bibitem{Ambrozewicz2007}
P.~Ambrozewicz et~al. (CLAS Collaboration), Phys. Rev. \textbf{C 75}, 045203
  (2007)

\bibitem{Coman2010}
M.~Coman et~al. (Jefferson Lab Hall A Collaboration), Phys. Rev. \textbf{C 81},
  052201 (2010)

\bibitem{McCracken2010}
M.E. McCracken et~al. (CLAS Collaboration), Phys. Rev. \textbf{C 81}, 025201
  (2010)

\bibitem{Blomqvist1998}
K.I. Blomqvist et~al. (A1 Collaboration), Nucl. Instr. and Meth. Phys. Res.
  \textbf{A 403}, 263 (1998)

\bibitem{Achenbach2011:EPJST}
P.~Achenbach et~al. (A1 Collaboration), Eur. Phys. J. Special Topics
  \textbf{198}, 307 (2011)

\bibitem{Achenbach2011:MWPC}
P.~Achenbach et~al. (A1 Collaboration), Nucl. Instr. Meth. Phys. Res. \textbf{A
  641}, 105 (2011)

\bibitem{MoTsai1969}
L.W. Mo, Y.S. Tsai, Rev. Mod. Phys. \textbf{41}, 205 (1969)

\bibitem{Glander2003}
K.H. Glander et~al. (SAPHIR Collaboration), Eur. Phys. J. \textbf{A 19}, 251
  (2004)

\bibitem{Niculescu1998}
G.~Niculescu et~al. (E93-018 Collaboration), Phys. Rev. Lett. \textbf{81}, 1805
  (1998)

\bibitem{MartAndTiator}
T.~Mart, L.~Tiator (2011), private communication, to be published.

\end{thebibliography}

\clearpage
\newpage
\onecolumn


%

\end{document}